\documentclass[prl,twocolumn,aps,superscriptaddress,showpacs,nofootinbib]{revtex4-1}
\usepackage{amsfonts}
\usepackage{amsmath}
\usepackage{amsthm}
\usepackage{amscd}
\usepackage{amssymb}
\usepackage{subfigure}
\usepackage{amsxtra}
\usepackage{gensymb}
\usepackage{bm}           
\usepackage{bbm}
\usepackage{graphicx}
\usepackage{epstopdf}
\usepackage{color}
\usepackage{times}
\usepackage{mdframed}
\usepackage{textpos}
\usepackage[colorlinks=true,citecolor=blue,urlcolor=black]{hyperref}

\def\bi#1\ei {\begin{itemize}#1\end{itemize}}
\def\bn#1\en {\begin{enumerate}#1\end{enumerate}}
\def\bea#1\eea {\begin{align}#1\end{align}}
\def\bean#1\eean {\begin{align*}#1\end{align*}}
\def\ben#1\een {\begin{equation*}#1\end{equation*}}
\def\be#1\ee {\begin{equation}#1\end{equation}}
\def\bes#1\ees {\begin{equation}\begin{split}#1\end{split}\end{equation}}
\def\bear#1\eear {\begin{eqnarray}#1\end{eqnarray}}
\def\bear#1\eear {\begin{eqnarray*}#1\end{eqnarray*}}

\newcommand{\beq}{\begin{equation}}
\newcommand{\eeq}{\end{equation}}

\begin{document}

\title{\bf Chromatic interferometry with small frequency differences}
\begin{textblock*}{5cm}(14.5cm,-1.5cm)
  \fbox{\footnotesize MIT-CTP-5111}
\end{textblock*}

\author{Luo-Yuan Qu}
\affiliation{Hefei National Laboratory for Physical Sciences at the Microscale and Department of Modern Physics, University of Science and Technology of China, Hefei 230026, China}
\affiliation{Shanghai Branch, CAS Center for Excellence in Quantum Information and Quantum Physics, University of Science and Technology of China, Shanghai 201315, China}
\affiliation{Shanghai Research Center for Quantum Sciences, Shanghai 201315, China}
\affiliation{Jinan Institute of Quantum Technology, Jinan, 250101, P.~R.~China}

\author{Lu-Chuan Liu}
\affiliation{Hefei National Laboratory for Physical Sciences at the Microscale and Department of Modern Physics, University of Science and Technology of China, Hefei 230026, China}
\affiliation{Shanghai Branch, CAS Center for Excellence in Quantum Information and Quantum Physics, University of Science and Technology of China, Shanghai 201315, China}
\affiliation{Shanghai Research Center for Quantum Sciences, Shanghai 201315, China}

\author{Jordan Cotler}
\affiliation{Society of Fellows, Harvard University, Cambridge, MA 02138 USA}
\affiliation{Stanford Institute for Theoretical Physics, Stanford University, Stanford, CA 94305 USA}

\author{Fei Ma}
\affiliation{Hefei National Laboratory for Physical Sciences at the Microscale and Department of Modern Physics, University of Science and Technology of China, Hefei 230026, China}
\affiliation{Shanghai Branch, CAS Center for Excellence in Quantum Information and Quantum Physics, University of Science and Technology of China, Shanghai 201315, China}
\affiliation{Shanghai Research Center for Quantum Sciences, Shanghai 201315, China}
\affiliation{Jinan Institute of Quantum Technology, Jinan, 250101, P.~R.~China}

\author{Jian-Yu Guan}
\affiliation{Shanghai Branch, National Laboratory for Physical Sciences at Microscale and Department of Modern Physics University of Science and Technology of China, Shanghai, 201315, P.~R.~China}
\affiliation{CAS Center for Excellence and Synergetic Innovation Center in Quantum Information and Quantum Physics, Shanghai Branch,  University of Science and Technology of China, Shanghai, 201315, P.~R.~China}

\author{Ming-Yang Zheng}
\author{Quan Yao}
\author{Xiu-Ping Xie}
\affiliation{Jinan Institute of Quantum Technology, Jinan, 250101, P.~R.~China}

\author{Yu-Ao Chen}
\author{Qiang Zhang}
\affiliation{Hefei National Laboratory for Physical Sciences at the Microscale and Department of Modern Physics, University of Science and Technology of China, Hefei 230026, China}
\affiliation{Shanghai Branch, CAS Center for Excellence in Quantum Information and Quantum Physics, University of Science and Technology of China, Shanghai 201315, China}
\affiliation{Shanghai Research Center for Quantum Sciences, Shanghai 201315, China}

\author{Frank Wilczek}
\affiliation{Center for Theoretical Physics, MIT, Cambridge, MA 02139 USA}
\affiliation{T. D. Lee Institute, Shanghai Jiao Tong University, Shanghai, 200240, P.~R.~China}
\affiliation{Wilczek Quantum Center, School of Physics and Astronomy, Shanghai Jiao Tong University, Shanghai, 200240, P.~R.~China}
\affiliation{Department of Physics, Stockholm University, Stockholm SE-106 91 Sweden}
\affiliation{Department of Physics and Origins Project, Arizona State University, Tempe, AZ 25287 USA}

\author{Jian-Wei Pan}
\affiliation{Hefei National Laboratory for Physical Sciences at the Microscale and Department of Modern Physics, University of Science and Technology of China, Hefei 230026, China}
\affiliation{Shanghai Branch, CAS Center for Excellence in Quantum Information and Quantum Physics, University of Science and Technology of China, Shanghai 201315, China}
\affiliation{Shanghai Research Center for Quantum Sciences, Shanghai 201315, China}


\begin{abstract}
By developing a `two-crystal' method for color erasure, we can broaden the scope of chromatic interferometry to include optical photons whose frequency difference falls outside of the 400 nm to 4500 nm wavelength range, which is the passband of a PPLN crystal. We demonstrate this possibility experimentally, by observing interference patterns between sources at 1064.4 nm and 1063.6 nm, corresponding to a frequency difference of about 200 GHz.
\end{abstract}

\maketitle

\section{Introduction}

Chromatic interferometry refers broadly to experiments which leverage quantum superposition in frequency-space to recover hidden phase information encoded in correlations among photons with different wavelengths~\cite{qu2019color}.  Recently, chromatic interferometry has attracted increasing attention, both for its intrinsic interest and for its possible utility in high-resolution imaging and photonic computation \cite{cotler2016entanglement, kobayashi2016frequency, kobayashi2017mach, lu2018electro, lu2018quantum, kues2019quantum, qu2019color}.


Color erasure is the essential technology enabling chromatic interferometry. Only when the frequency difference between the photons is surpassed by the response of detector can interference be measured\cite{vittorini2014entanglement,guo2017testing,wang2018experimental}. Information which identifies wavelength (e.g., specifically, energy deposit) is registered in the detection apparatus, even if it is not readily accessible to an experimentalist.  Wavelength information is generally harder to erase than polarization or path information, and so color erasure poses an interesting challenge.

The purpose of ``color erasure detectors''\cite{cotler2016entanglement, qu2019color} is to erase all wavelength identifying information, thus enabling chromatic interference.  The use of such detectors goes beyond previous experiments in chromatic interferometry which implement wavelength conversion either at the light source, or at beamsplitters \cite{takesue2008erasing, raymer2010interference, de2012quantum, kobayashi2016frequency, kobayashi2017mach, lu2018quantum}.  By contrast, color erasure detectors can recover phase information between different wavelengths of light \textit{after} interference or phase accumulation has occurred.

Ironically, a significant limitation of existing color erasure detectors is that they can only render photons indistinguishable when their frequency difference is sufficiently large. In order to render reception of two optical photons with frequencies $f_1 <  f_2$ indistinguishable, they employ three-wave mixing with a coherent source at frequency $f_3 = f_2 - f_1$.  Appropriate crystals or waveguides that implement the mixing are available if $f_3$ corresponds to a wavelength in the $400$ nm to $4500$ nm wavelength range, but not otherwise.  This consideration significantly restricts the frequencies $f_1$ and $f_2$ and thus the scope of applications.

Here we develop a more general method of color erasure, which allows $f_2 - f_1$ to be very small.  We demonstrate its soundness and practicality by performing chromatic intensity (Hanbury Brown--Twiss) interferometry~\cite{brown1956test, twiss1957question} between sources with 1064.4 nm and 1063.6 nm photons.  Hanbury Brown--Twiss interferometry plays an important role in quantum optics \cite{scully1999quantum} and has wide applications in astronomy and fluorescence microscopy \cite{brown1956test,monnier2003optical, schwartz2013superresolution, grussmayer2014photon}, and so our experiments lay the groundwork for new chromatic generalizations and technologies.

\section{Theory}

Our goal is to develop a detector that cannot distinguish between photons with optical frequencies $f_1$ and $f_2$.  We introduce a third frequency $f_3$ with $f_1 < f_2 < f_3$ and such that $\Delta f_{31} = f_3 - f_1$ and $\Delta f_{32} = f_3 - f_2$ are both optical frequencies.  Denote photons of frequency $f_1, f_2, f_3$ by $\gamma_1, \gamma_2, \gamma_3$, and photons with frequency $f_1' = f_1 + \Delta f_{32}$ and $f_2' = f_2 + \Delta f_{31}$ by $\gamma_1'$ and $\gamma_2'$ respectively.

Let us first describe our protocol heuristically, to provide intuition for the mathematics to follow.   Consider a superposition of photons with wavelengths $f_1$, $f_2$.  Using a beamsplitter, we can transform this state into a (further) superposition of two distinct spatiotemporal modes.  Let us put the photons in the first mode through a PPLN waveguide \cite{ma20171} pumped with a coherent state of many $\Delta f_{31}$ photons.  In this way, we induce upconversions  $f_1 \to f_3$ and $f_2 \to f_2'$.  Similarly, let us put photons in the second mode through a second PPLN waveguide pumped with a coherent state of many $\Delta f_{21}$ photons, inducing upconversions $f_1 \to f_1'$ and $f_2 \to f_3$.   Then we can filter both beams to allow only photons with frequency $f_3$, and finally recombine the two beams using a second beamsplitter.  This processing and filtering renders it impossible to determine whether the triggering photons had frequency $f_1$ or $f_2$.

Now let us treat this mathematically.  Let $|\Omega\rangle$ be the vacuum state, and let $a_\gamma^\dagger$ create a $\gamma$ photon in some fixed spatiotemporal mode.  Then, for instance, $a_\gamma^\dagger a_{\gamma'}^\dagger$ would create two photons $\gamma$ and $\gamma'$ in the same fixed spatiotemporal mode.  For simplicity, consider the initial state
\begin{equation}
|\Psi_0\rangle = \left(\alpha \,a_{\gamma_1}^\dagger+ \beta \,a_{\gamma_2}^\dagger\right) |\Omega\rangle
\end{equation}
where $|\alpha|^2 + |\beta|^2 = 1$.  This state corresponds to a superposition of a $\gamma_1$ photon and a $\gamma_2$ photon in a single spatiotemporal mode.

Consider a second spatiotemporal mode, with corresponding creation operators given by $b_\gamma^\dagger$\,.  A 50-50 beamsplitter between the first and second spatiotemporal modes corresponds to
\begin{equation}
a_\gamma^\dagger \longrightarrow \frac{1}{\sqrt{2}}\left( a_\gamma^\dagger + b_\gamma^\dagger \right)\,, \qquad b_\gamma^\dagger \longrightarrow \frac{1}{\sqrt{2}}\left( a_\gamma^\dagger - b_\gamma^\dagger \right)
\end{equation}
for all $\gamma$.  Applying such a 50-50 beamsplitter to $|\Psi_0\rangle$, we obtain
\begin{equation}
\label{eq:twocrystalsplitstep1}
\frac{1}{\sqrt{2}}\left[\left(\alpha \,a_{\gamma_1}^\dagger+ \beta \,a_{\gamma_2}^\dagger\right) + \left(\alpha \,b_{\gamma_1}^\dagger+ \beta \,b_{\gamma_2}^\dagger\right) \right] |\Omega\rangle\,.
\end{equation}

Evolution of the first (second) mode, propagating through a PPLN waveguide pumped with a coherent state of a large $N$ number of $\Delta f_{31}$ ($\Delta f_{32}$) photons, is described by the Hamiltonian $H_{31}$ ($H_{32}$) where 
\begin{align}
H_{31} &= i \,\xi_{31} \left(e^{i \phi_{31}} \, a_{\gamma_1} a_{\gamma_3}^\dagger - e^{- i \phi_{31}} a_{\gamma_1}^\dagger a_{\gamma_3} \right) \nonumber \\
& \qquad \quad + i \,\xi_{2'2} \left(e^{i \phi_{2'2}} \, a_{\gamma_2} a_{\gamma_2'}^\dagger - e^{- i \phi_{2'2}} a_{\gamma_2}^\dagger a_{\gamma_2'} \right) \\
H_{32} &= i \,\xi_{32} \left(e^{i \phi_{32}} \, b_{\gamma_2} b_{\gamma_3}^\dagger - e^{- i \phi_{32}} b_{\gamma_2}^\dagger b_{\gamma_3} \right) \nonumber \\
& \qquad \quad + i \,\xi_{1'1} \left(e^{i \phi_{1'1}} \, b_{\gamma_1} b_{\gamma_1'}^\dagger - e^{- i \phi_{1'1}} b_{\gamma_1}^\dagger b_{\gamma_1'} \right) \,.
\end{align}
The $\xi$ parameters control the speed of up- and down-conversion, and the $\phi$ parameters dictate the phases accumulated by the converted photons during the process.  These effective Hamiltonians, which cause the $f_1$ and $f_2$ photons to become entangled with the large $N$ coherent state of the pump, were derived in \cite{qu2019color} using a systematic $1/N$ expansion.  As emphasized in \cite{qu2019color}, large $N$ coherent states are physically essential, since we want to `lose track' of the loss or gain of single photons.  In our setup, we consider
the combined Hamiltonian
\begin{equation}
H = H_{31} + H_{32}
\end{equation}
and evolve \eqref{eq:twocrystalsplitstep1} by $e^{-i H T}$.

We apply a second 50-50 beamsplitter to both spatiotemporal modes, and finally filter to $\gamma_3$ photons in the first outputted spatiotemporal mode.  This corresponds to projecting onto $a_{\gamma_3}^\dagger |\Omega\rangle$.  The resulting state is
\begin{align}
\frac{1}{2}\left(\alpha \, e^{i \phi_{31}} \sin(\theta_{31}) + \beta \, e^{i \phi_{32}}\cos(\theta_{32}) \right) a_{\gamma_3}^\dagger |\Omega\rangle\,
\end{align}
where $\theta_{ij} \equiv T \xi_{ij}$.  These angular $\theta_{ij}$ parameters control the amount of up- and down-conversion that have occurred between the photons with frequencies $f_i$ and $f_j$.  By tuning $\phi_{31} = \phi_{32} = 0$, and say $\theta_{31} = \pi/2$ and $\theta_{32} = 2\pi$, we get
\begin{align}
\frac{1}{2}\left(\alpha  + \beta \right) a_{\gamma_3}^\dagger|\Omega\rangle\,
\end{align}
Putting everything together, we have
\begin{equation}
\label{eq:totaldynamics}
\left(\alpha \,a_{\gamma_1}^\dagger+ \beta \,a_{\gamma_2}^\dagger\right) |\Omega\rangle \,\, \longrightarrow \,\, \frac{1}{2}\left(\alpha  + \beta \right) a_{\gamma_3}^\dagger |\Omega\rangle
\end{equation}
as was desired.  What we have effectively done is mapped $a_{\gamma_1}^\dagger \to \frac{1}{\sqrt{2}} \, a_{\gamma_3}^\dagger + \cdots$ and $a_{\gamma_2}^\dagger \to \frac{1}{\sqrt{2}} \, a_{\gamma_3}^\dagger + \cdots$, and then post-selected onto the outcome of receiving a $\gamma_3$ photon.

Our arrangement in its entirety embodies a single color erasure detector.  Equation~\eqref{eq:totaldynamics} summarizes the manner in which the detector decoheres a state \cite{zurek2003decoherence,zurek2009quantum}.  Color erasure is achieved through an entangling measurement, as described above.

We conclude this section by seeing how color erasure detectors allow us to perform Hanbury Brown-Twiss interferometry with sources having distinct wavelength.  Here we will be schematic, but full details can be found in~\cite{qu2019color, cotler2016entanglement}.  Suppose we consider the standard Hanbury Brown-Twiss experiment with two sources of the \textit{same} wavelength.  Let $a_{\gamma}^\dagger$ and $b_{\gamma}^\dagger$ denote creation operators for $\gamma$ photons at the locations of two detectors $A$ and $B$, respectively.  Suppose each source emits a single photon at some moment in time.  Then once the photons have reached the detectors, we will have a state
\begin{equation}
\label{eq:HBTschematic1}
(\alpha + \beta) a_{\gamma}^\dagger b_{\gamma}^\dagger |\Omega\rangle + [\text{orthogonal states}]
\end{equation}
which allows us to extract $|\alpha + \beta|^2 = |\alpha|^2 + \alpha \beta^* + \alpha^* \beta + |\beta|^2$ corresponding to the probability that each detector received exactly one of the two photons (i.e., a coincidence count).  This probability crucially contains an interference term $\alpha \beta^* + \alpha^* \beta$, which encodes desired phase information in the Hanbury Brown-Twiss setup.

By contrast, if the first source emits photons of wavelength $\gamma_1$ and the second source emits photons of wavelength $\gamma_2$, then the analog of Eqn.~\eqref{eq:HBTschematic1} is
\begin{equation}
\label{eq:HBTschematic2}
(\alpha \, a_{\gamma_1}^\dagger b_{\gamma_2}^\dagger + \beta \, a_{\gamma_2}^\dagger b_{\gamma_1}^\dagger) |\Omega\rangle + [\text{orthogonal states}]\,.
\end{equation}
Since $a_{\gamma_1}^\dagger b_{\gamma_2}^\dagger |\Omega\rangle$ and $a_{\gamma_2}^\dagger b_{\gamma_1}^\dagger |\Omega\rangle$ are orthogonal, we can only extract $|\alpha|^2$ and $|\beta|^2$ via measurement, and so we do not have access to the interference term $\alpha \beta^* + \alpha^* \beta$.  To gain access to this interference term, we can let detectors $A$ and $B$ be color erasure detectors, taking $a_{\gamma_1}^\dagger b_{\gamma_2}^\dagger |\Omega\rangle \to \frac{1}{4}\,a_{\gamma_3}^\dagger b_{\gamma_3}^\dagger |\Omega\rangle$  and $a_{\gamma_2}^\dagger b_{\gamma_1}^\dagger |\Omega\rangle \to \frac{1}{4}\,a_{\gamma_3}^\dagger b_{\gamma_3}^\dagger |\Omega\rangle$ as per~\eqref{eq:totaldynamics}.  Accordingly,~\eqref{eq:HBTschematic2} becomes
\begin{equation}
\label{eq:HBTschematic3}
\frac{1}{4}(\alpha + \beta) a_{\gamma_3}^\dagger b_{\gamma_3}^\dagger |\Omega\rangle + [\text{orthogonal states}]
\end{equation}
from which we can extract $|\alpha + \beta|^2$ and the desirable interference term $\alpha \beta^* + \alpha^* \beta$ by determining the frequency of coincidence counts of $\gamma_3$ photons at the color erasure detectors $A$ and $B$.

In summary, color erasure detectors allow us to perform Hanbury Brown-Twiss interferometry using the standard procedure, even when the sources have distinct wavelength.  We will experimentally implement this color erasure version of Hanbury Brown-Twiss interferometry in the next section.

\section{Experiment}
We have implemented the theoretical proposal given above and used the resulting detectors to perform chromatic intensity interferometry. As shown in Figure 1, 1064.4 nm photons and  1063.6 nm photons, prepared in weak coherent states, meet at a 50-50 beamsplitter labeled BS1. The linewidth of the photons is about 1 kHz. The 1064.4 nm photons and 1063.6 nm photons will not mutually interfere because their frequency difference is about 200 GHz.  To recover chromatic interference, we build up two color erasure detectors, each having a traditional Si single photon detector, two beamsplitters, a special-made PPLN waveguide \cite{ma20171}, a pump laser, and a filter.

\begin{figure*}[tbh]
\centering
\resizebox{14cm}{!}{\includegraphics{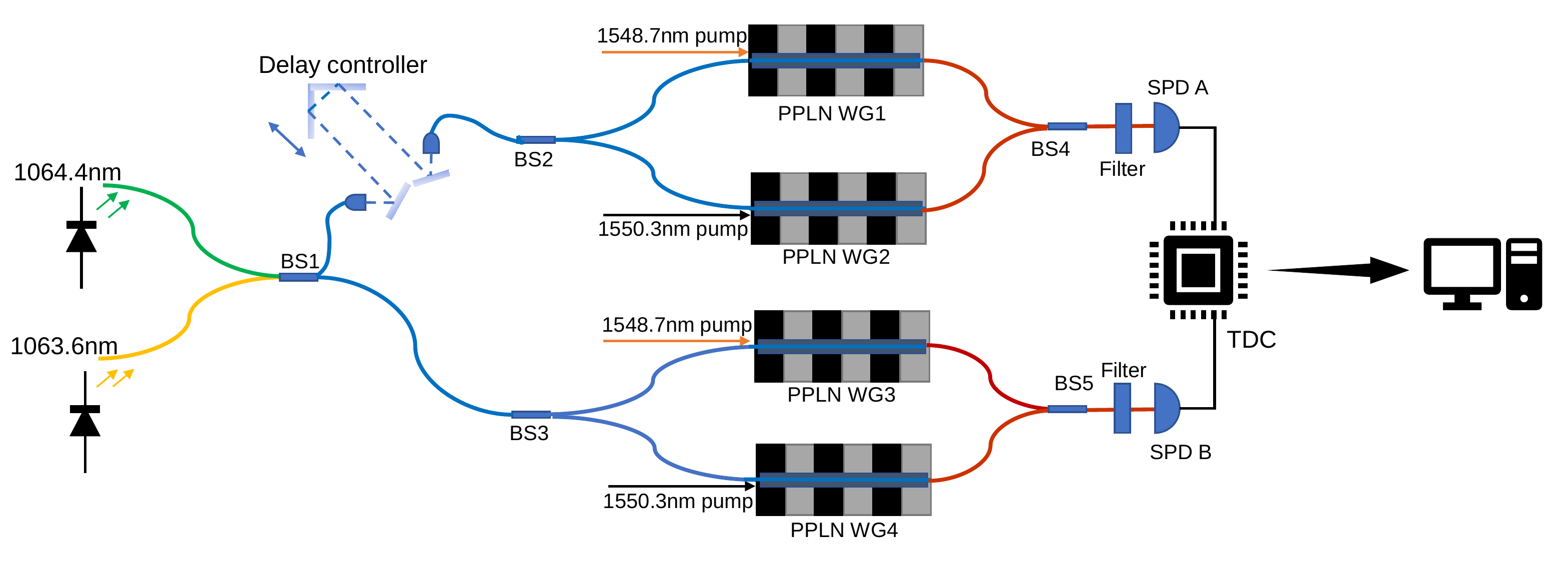}}
\caption{\textbf{Diagram of the intensity interferometer.} Abbreviations are: periodically-poled lithium niobate waveguide (PPLN WG), beamsplitter (BS), single photon detector (SPD), time-to--digital converter (TDC).}
\label{fig1}
\end{figure*}

Figure~\ref{fig1} shows a diagram of the setup.  After the initial BS1, the superposed mixture of 1064.4 nm photons and 1063.6 nm photons is further split and superposed by BS2 and BS3. A delay controller is inserted before BS2 to control the phase of the photons. The two paths emanating from BS2 go into two separate PPLN waveguides (denoted by PPLN WG 1 and 2), and similarly the two paths emanating from BS3 go into separate waveguides (labeled PPLN WG 3 and 4).  We use a 1548.7 nm laser with about 100 kHz linewidth to pump PPLN WG 1 and 3, and a 1550.3 nm laser with about 1 kHz linewidth to pump PPLN WG 2 and 4. The outputs of PPLN WG 1 and 2 are coupled to single photon detector A (SPD A) by BS4 and the outputs of PPLN WG 3 and 4 are coupled to single photon detector B (SPD B) by BS5. In PPLN WG 1 and 3, we convert 1064.4 nm photons to 630.8 nm photons via sum-frequency generation (SFG). In PPLN WG 2 and 4, we convert 1063.6 nm photons to 630.8 nm photons via SFG.  A 630.8 nm filter allows us to filter in only the 630.8 nm photons.

When SPD A or SPD B receives a 630.8 nm photon, it in principle cannot tell if the photon was originally 1064.4 nm or 1063.6 nm.  Every photon arrival time at SPD A and SPD B is recorded by a time-to-digital converter (TDC).  To observe chromatic Hanbury Brown--Twiss interferometry, we measure the $g^{(2)}$ correlation. Our calculation of $g^{(2)}(\tau)$ amounts to
\begin{equation}
g^{(2)}(\tau)=\frac{n_{\text{coincidence}} \cdot n_{\text{bin}}}{n_{A} \cdot n_{B}}
\end{equation}
where $n_{\text{coincidence}}$ is the number of coincidence counts between SPD A and SPD B, $n_{\text{bin}}$ is the number of time bins in our trial, $n_A$ is the number of counts at SPD A, and $n_B$ is the number of counts at SPD B.
Also, $\tau$ is the delay applied on the signal of detector B.

The second order correlation $g^{(2)}(\tau = 0)$ of two lasers is (see the supplemental materials of~\cite{qu2019color}, as well as~\cite{cotler2016entanglement}, for a detailed derivation)
\begin{equation}
g^{(2)}(\tau = 0)=1+\frac{\varepsilon }{2}\,\cos(\Delta\phi_{1AB} - \Delta\phi_{2AB})
\end{equation}
where $\varepsilon$ is the visibility of the interferometry.  Above, $\Delta \phi_{1AB}$ is the phase difference between the paths from the first source to A and the first source to B, whereas $\Delta \phi_{2AB}$ is the phase difference between the paths from the second source to A and the second source to B. In our experimental color erasure setting, the phases from the sources to detector B are fixed.  By adjusting the reflector in the delay controller, we can increase the optical path by $\Delta L$, and thus $t_{delay}=\frac{\Delta L}{c}$ and $\Delta \phi =2\pi\Delta f_{21} \, t_{delay}$.  Then we can write $g^{(2)}(\tau = 0)$ more explicitly as \cite{cotler2016entanglement, qu2019color}
\begin{equation}
\label{E:g2eq1}
g^{(2)}(\tau = 0)=1+\frac{\varepsilon}{2}\, \cos(\phi_{0} + 2\pi\Delta f_{21} \, t_{delay})\,.
\end{equation}
We can also write a more explicit expression for $\varepsilon$.  Let $n_{1A}$ be the number of photons from source 1 which arrive at detector A, and similarly define $n_{2A}$, $n_{1B}$, $n_{2B}$.  We also let $n_{dA}$ and $n_{dB}$ denote the unwanted photon counts, including dark counts, environment light, and counts from unfiltered signal and pump light.  Then we have
\begin{equation}
\varepsilon=\frac{4\sqrt{n_{1A}n_{2A}n_{1B}n_{2B}}}{(n_{1A}+n_{2A}+n_{dA})(n_{1B}+n_{2B}+n_{dB})}\,.
\end{equation}

\begin{figure}[t]
\centering
\includegraphics[width=0.42\textwidth]{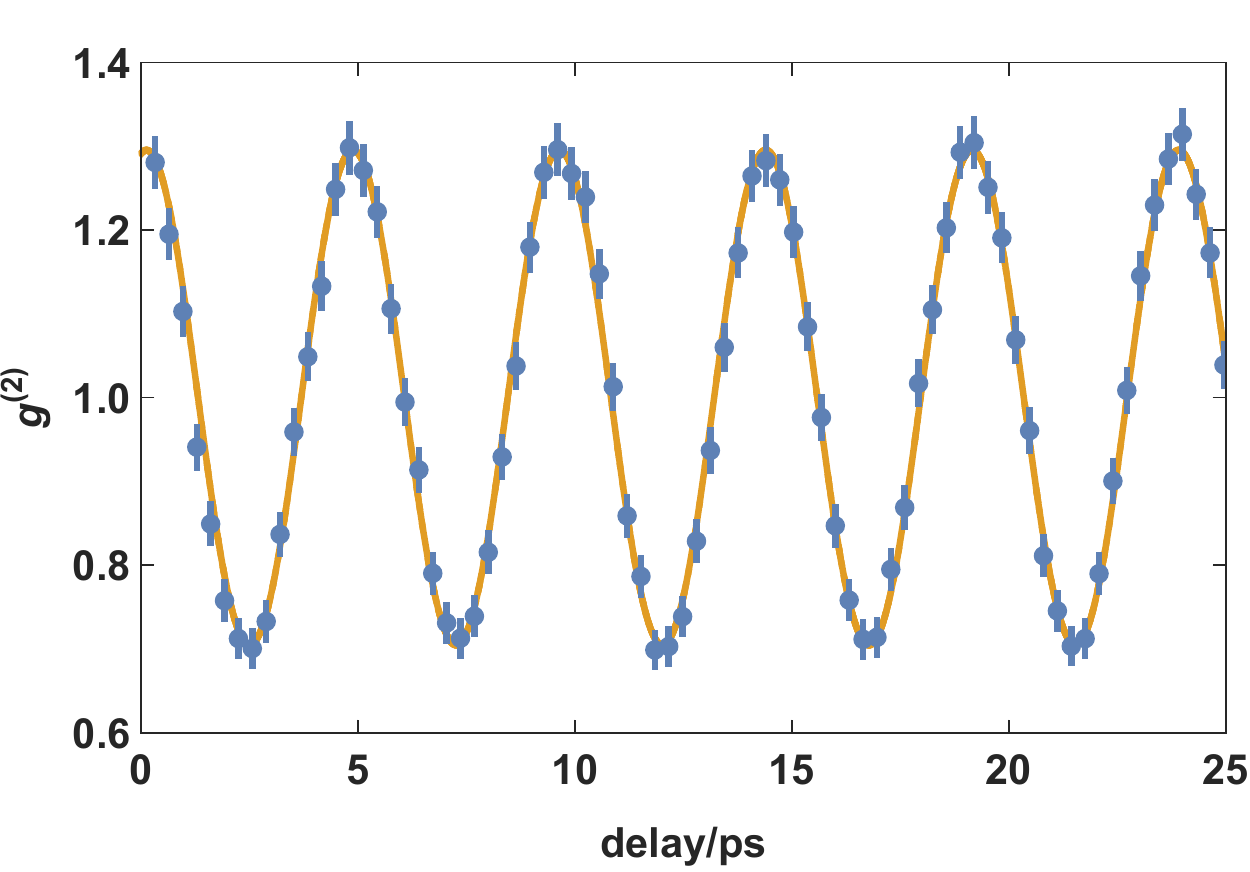}
\caption{\textbf{Intensity interferometry of two lasers.} The blue dots represent $g^{(2)}(\tau =0)$ with different phases controlled by the delay controller. The yellow line is the fitting result, which encodes the frequency difference between the 1064.4 nm and 1063.6 nm photons.}
\label{fig2}
\end{figure}

As shown in Figure~\ref{fig2}, $g^{(2)}(\tau = 0)$ oscillates as we change the phase of the interferometer.  If the photons were still distinguishable upon measurement, we would have $g^{(2)}(\tau = 0) = 1$.  Instead, since the color erasure detectors render the photons indistinguishable, $g^{(2)}(\tau = 0)$ need not be near one.

As we see in Figure~\ref{fig2}, by changing the length of the optical path from the output of BS1 to detector A, the photons can both bunch and anti-bunch when they arrive at the detectors \cite{hong1987measurement}.  Performing a least squares fitting to~\eqref{E:g2eq1}, we find
\begin{equation}
\label{E:params1}
\begin{aligned}
\varepsilon&=0.59\pm 0.01 \\
\phi_{0}&=-0.16\pm0.04 \\
\Delta f_{21}&=210.1\pm0.5 \,\text{GHz}\\
\end{aligned}
\end{equation}
This is consistent with our experimental parameters since the frequency difference between $1064.4$ nm and $1063.6$ nm corresponds to $\approx 212$ GHz (with some systematic uncertainty corresponding to drifting of the sources by up to several GHZ around $212$ GHz).

\begin{figure}[t]
\centering
\includegraphics[width=0.42\textwidth]{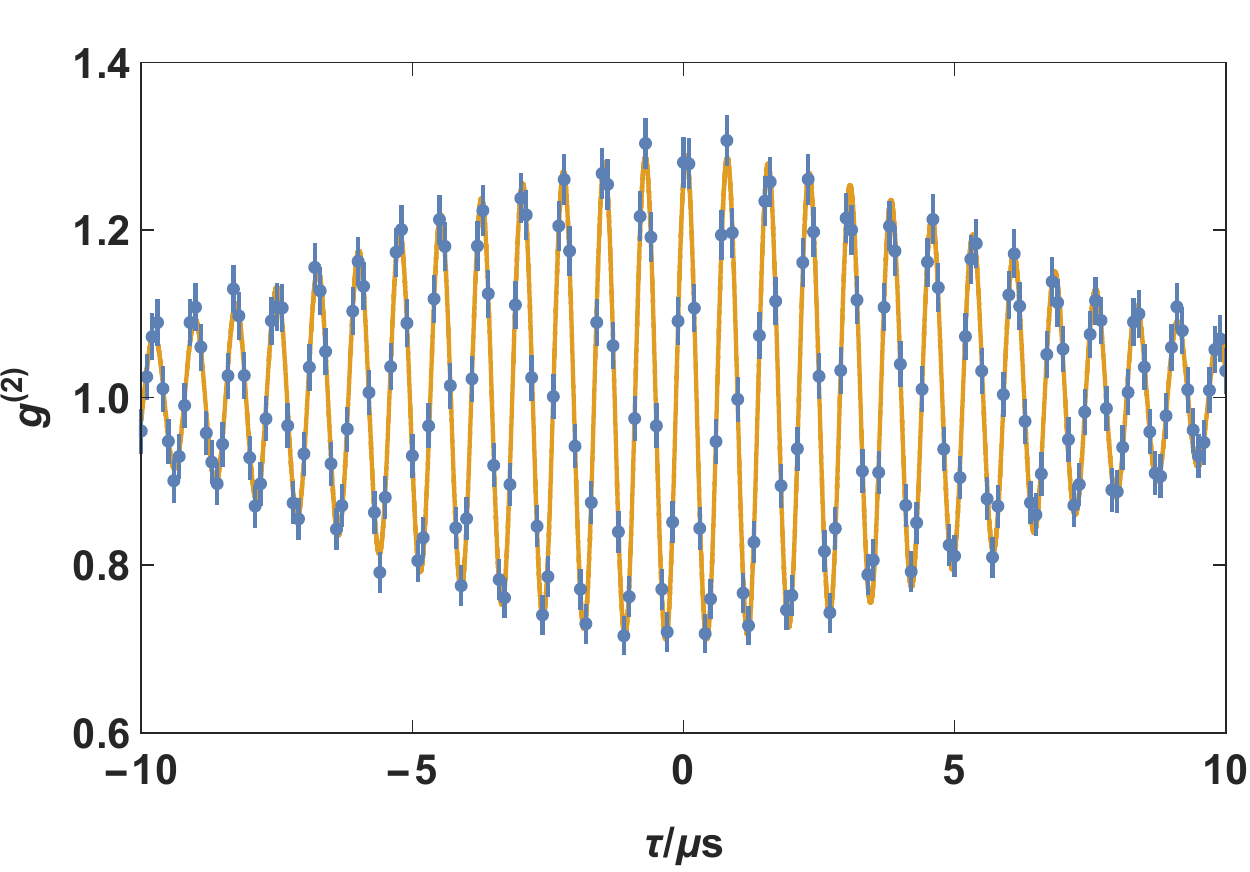}
\caption{\textbf{$g^{(2)}(\tau)$ at different $\tau$.} The blue dots are calculated from data with a fixed time delay from the delay controller, and $\tau$ added to the timestamp of the SPD B detector in post-processing. The yellow line is the fitting result, which encodes the original frequency difference of the color-erased photons, as well as their coherence.} 
\label{fig3}
\end{figure}

The frequencies of the pump lasers are carefully tuned so that the received photons are indistinguishable to the Si APD.  Although ideally we are engineering the processes $f_1 \to f_3$ and $f_2 \to f_3$, in reality we have $f_1 \to f_3^{(1)}$ and $f_2 \to f_3^{(2)}$ where $f_3^{(1)} \approx f_3^{(2)}$.  This is okay, so long as $f_3^{(1)}$ and $f_3^{(1)}$ are close enough to be rendered indistinguishable due to the time resolution of the receiving detectors.  The difference $f_3^{(2)} - f_3^{(1)}$ appears in the theoretical formula for $g^{(2)}(\tau)$, namely
\begin{equation}
\label{E:g2eq2}
g^{(2)}(\tau) = 1 + \frac{\varepsilon}{2} \, e^{- \gamma^2 \tau^2} \cos(\phi_1 + 2 \pi |f_3^{(2)} - f_3^{(1)}| \tau )\,,
\end{equation}
where $\gamma$ is the spectral linewidth.

Indeed, as shown in Figure~\ref{fig3}, $g^{(2)}(\tau)$ oscillates as we apply different $\tau$ by post-processing. The speed of the oscillations encodes the original frequency difference of the color-erased photons, and this is not faster than the time resolution of the detectors since otherwise the observed interference would vanish.  Also, the interference decays as $\tau$ surpasses the coherence time of the detected photons.  A least squares fitting to~\eqref{E:g2eq2} gives
\begin{equation}
\begin{aligned}
\varepsilon&=0.576 \pm 0.008 \\
\gamma &= 0.118 \pm 0.002 \, \text{MHz}\\
\phi_{1}&=-0.434 \pm 0.011 \\
|f_3^{(2)} - f_{3}^{(1)}| &= 1.32 \pm 0.02 \,\text{MHz}\\
\end{aligned}
\end{equation}
The fitted value of $\varepsilon$ for $g^{(2)}(\tau)$ is necessarily similar to the fitted value for $\varepsilon$ for $g^{(2)}(\tau = 0)$ in~\eqref{E:params1}, and the fitted value of the spectral linewidth $\gamma$ is consistent with known experimental parameters.

\section{Discussion}

We have presented a new methodology for color erasure detectors which enables chromatic interferometry of photons with small frequency differences.  This more general method can also be used for large frequency differences, as an alternative to the procedure in \cite{qu2019color}.


Multi-photon interference enables higher phase sensitivity to light sources, and better resolution of their geometries.  However, if the source or sources in question emit photons with distinct wavelengths, then interference between their emitted photons will not occur and the desired phases cannot be extracted.  But color erasure detectors allow one to gain access to the desired phase information by retroactively recovering interference (akin to a quantum eraser~\cite{scully1991quantum, kwiat1992observation}) between the photons emitted from the sources.

In several circumstances, including stars or exoplanets \cite{monnier2003optical} having very different temperatures or differentially fluorescent structures\cite{shtengel2009interferometric, leung2011review, schwartz2013superresolution, grussmayer2014photon}, chromatic interferometry promises to be a natural tool for achieving high resolution.
We are actively pursuing these directions. \\

\section*{Funding}
National Key R\&D Program of China (No.2018YFB0504300); the National Natural Science Foundation of China; the Chinese Academy of Sciences (CAS);Shanghai Municipal Science and Technology Major Project (Grant No.2019SHZDZX);  Anhui Initiative in Quantum Information Technologies; Junior Fellowship from the Harvard Society of Fellows; Fannie and John Hertz Foundation; the Stanford Graduate Fellowship program; the U.S. Department of Energy under grant Contract  Number DE-SC0012567; the European Research Council under grant 742104; Swedish Research Council under Contract No. 335-2014-7424. 

\section*{Acknowledgments}
We thank Hai-feng Jiang and Qi Shen for their experimental assistance.


\bibliography{Bibliography.bib}


  







\newpage
\end{document}